\pdfoutput=1
\documentclass[conference]{IEEEtran}
\IEEEoverridecommandlockouts
\usepackage{cite}
\usepackage{csquotes}
\usepackage{amsmath,amssymb,amsfonts}
\usepackage{algorithmic}
\usepackage{graphicx}
\usepackage{textcomp}
\usepackage{xcolor}
\usepackage{xspace}
\usepackage{hyperref}
\usepackage{flushend}
\usepackage{tikz}
\usetikzlibrary{svg.path}
    
\usepackage[activate={true,nocompatibility},final,tracking=true,kerning=true,spacing=true]{microtype}
\SetTracking{encoding=*, shape=sc}{30} 
    
\newtheorem{definition}{Definition}

\newcommand\copyrighttext{%
  \footnotesize \textcopyright~2021 IEEE. Personal use of this material is permitted.
  Permission from IEEE must be obtained for all other uses, in any current or future
  media, including reprinting/republishing this material for advertising or promotional
  purposes, creating new collective works, for resale or redistribution to servers or
  lists, or reuse of any copyrighted component of this work in other works.
  }
\newcommand\copyrightnotice{%
\begin{tikzpicture}[remember picture,overlay]
\node[anchor=south,yshift=20pt] at (current page.south) {\fbox{\parbox{\dimexpr\textwidth-\fboxsep-\fboxrule\relax}{\copyrighttext}}};
\end{tikzpicture}%
}

\newcommand{\ET}{\textsc{ET}\xspace}
    
\begin{document}

\title{On the Relation of Trust and Explainability:\\ Why to Engineer for Trustworthiness}

\author{\IEEEauthorblockN{
Lena Kästner\IEEEauthorrefmark{1},
Markus Langer \IEEEauthorrefmark{2},
Veronika Lazar\IEEEauthorrefmark{2},
Astrid Schomäcker\IEEEauthorrefmark{1},
Timo Speith\IEEEauthorrefmark{1}\IEEEauthorrefmark{3},
Sarah Sterz\IEEEauthorrefmark{1}\IEEEauthorrefmark{3}}
\IEEEauthorblockA{\IEEEauthorrefmark{1}Saarland University, Institute of Philosophy, Saarbrücken, Germany}
\IEEEauthorblockA{\IEEEauthorrefmark{2}Saarland University, Department of Psychology, Saarbrücken, Germany}
\IEEEauthorblockA{\IEEEauthorrefmark{3}Saarland University, Department of Computer Science, Saarbrücken, Germany}

Email: \{lena.kaestner, astrid.schomaecker, firstname.lastname\}@uni-saarland.de,  sterz@depend.uni-saarland.de
}

\maketitle

\copyrightnotice
\vspace{-2ex}

\begin{abstract}
Recently, requirements for the explainability of software systems have gained prominence. One of the primary motivators for such requirements is that explainability is expected to facilitate stakeholders' trust in a system. Although this seems intuitively appealing, recent psychological studies indicate that explanations do not necessarily facilitate trust. Thus, explainability requirements might not be suitable for promoting trust. 

One way to accommodate this finding is, we suggest, to focus on \emph{trustworthiness} instead of trust. While these two may come apart, we ideally want both: a trustworthy system and the stakeholder's trust. In this paper, we argue that even though trustworthiness does not automatically lead to trust, there are several reasons to engineer primarily for trustworthiness -- and that a system's explainability can crucially contribute to its trustworthiness.
\end{abstract}

\begin{IEEEkeywords}
Explainability, XAI, Trust, Trustworthiness, Requirements, NFR
\end{IEEEkeywords}

\section{Introduction}

Software systems used for decision-making are becoming increasingly complex and opaque. At the same time, such systems are used in processes of high social relevance, such as loan applications or parole decisions. It is an urgent question whether we should really \emph{trust} such opaque systems, which evade the understanding even of their programmers, to make critical decisions \cite{Panesar2019Ethics}. The concept of trust also plays an essential role in requirements engineering (RE), for instance, in ISO/IEC 25022 \cite{ISO25022}. However, trust remains a rather vague concept that is hard to measure and is, therefore, a difficult requirement to engineer towards \cite{ISO29148, Amershi2019Guidelines}.

Many see explainability as a suitable means to foster stakeholder trust \cite{Langer2021What, Chazette2021Exploring}: If we better understand how the system produces its outputs and the explanation for a given output fits with our expectations of how a good decision should be made, this explanation presents a reason to trust the system. Thus, at first glance, a requirement for explainability seems to be more suitable than to have a requirement for trust directly.

Its assumed potential to increase trust is one of the reasons why explainability has become a \enquote*{hot} topic in computer science and interdisciplinary research \cite{Langer2021What}, and now proliferates in the RE community as a non-functional requirement \cite{Chazette2021Exploring, Chazette2020Explainability, Koehl2019Explainability}. Indeed, explainability and trust are often connected in the literature \cite{
%
Langer2021What, Koehl2019Explainability, Chazette2020Explainability, Chazette2021Exploring,
%
Gilpin2018Explaining, Richardson2018Survey, Anjomshoae2019Explainable, Fox2017Explainable, Anjomshoae2019Explanations, Nalepa2018From, Atzmueller2019Towards, Paez2019Pragmatic, Pieters2011Explanation, Gregor1999Explanations, Hois2019How, Dam2018Explainable, 
%
Clinciu2019Survey, Cai2019Effects, Hoffman2018Explaining, Mathews2019Explainable, Wang2019Designing, DeGraaf2017How,
%
Abdul2018Trends, Adadi2018Peeking, Baaj2019Some, Balog2019Transparent, Baum2018From, Baum2018Towards, Carvalho2019Machine, Clos2017Towards, Conati2021Toward, Cotter2017Explainaing, Dodge2019Explaining, Freitas2014Comprehensible, Gilpin2018Society, Glass2008Toward, Green2009Generating, Guidotti2019Survey, Henin2019Towards, Holzinger2019Causability, Lage2019Exploring, Madumal2019Explainable, Madumal2019Grounded, Michael2019Machine, Miller2019Explanation, Nothdurft2013Impact, Olson2019Counterfactual, Ras2018Explanation, Ribeiro2016Why, Riedl2019Human, Rosenfeld2019Explainability, Sato2019Context, Schneider2019Personalized, Sevastjanova2018Going, Sheh2017Did, Sheh2018Defining, Sokol2018Conversational, Sokol2020Explainability, Sreedharan2018Handling, Swartout1983Xplain, Tintarev2007Explanations, Tintarev2007Effective, Tintarev2011Designing, Ter2017News, Vig2009Tagsplanations, Wang2018Explainable, Watts2019Local, Weber2019Explaining
} and many researchers, at least implicitly, assume some form of what we will call the \emph{Explainability-Trust-Hypothesis} (\ET) in the following:

\begin{description}
\item[(\ET)] Explainability is a suitable means for facilitating trust in a stakeholder.
\end{description}

Recent psychological research has shown, however, that this widely accepted hypothesis should be called into question. Several studies have shown either no effect or even a negative effect of explanations on subjects' trust in a system \cite{Chen2019User, Cheng2019Explaining, Kizilcec2016How, Papenmeier2019How}. In this paper, we will discuss what these findings tell us about the relationship between explainability and trust and how to proceed when engineering for trust based on explainability.

\section{\ET in the Literature}
The idea of a close connection between explanations or explainability and increased trust as expressed by \ET is pervasive in the literature on explainable AI (XAI). For illustration, consider the following quotes:
\begin{itemize}
    \item \enquote{In order for humans to trust black-box methods, we need explainability […].} \cite{Gilpin2018Explaining}
    \item \enquote{[…] in many, if not most, cases, the explanation is beneficial […] to foster better trust […].} \cite{Richardson2018Survey}
    \item \enquote{Increasing user’s trust in the system [… is] among the listed motivations for the explanations.} \cite{Anjomshoae2019Explainable}
    \item \enquote{The need for explainable AI is motivated mainly by three reasons: the need for trust […].} \cite{Fox2017Explainable}
    \item \enquote{Explanations are particularly essential […] as it [sic] raises trust […] in the system.} \cite{Anjomshoae2019Explanations}
    \item \enquote{[…] explainability will also enhance trust at the user side […].} \cite{Nalepa2018From}
    \item \enquote{[…] the provided […] explainability will also enhance trust in the system at the level of the users […].} \cite{Atzmueller2019Towards}
    \item \enquote{The main goal of Explainable Artificial Intelligence (XAI) has been variously described as as a search for explainability, […] for ways of […] generating trust in the model and its predictive performance.} \cite{Paez2019Pragmatic}
    \item \enquote{Artificial agents need to explain their decision to the user in order to gain trust […].} \cite{Pieters2011Explanation}
    \item \enquote{Explanations, by virtue of making the performance of a system transparent to its users, are influential […] for improving users' trust […].} \cite{Gregor1999Explanations}
    \item \enquote{[…] explainability provides transparency and contri\-butes to trust […].} \cite{Hois2019How}
    \item \enquote{Explainability is […] a pre-requisite for practitioner trust […].} \cite{Dam2018Explainable}
\end{itemize}

Other authors are more cautious. While they do connect explanations and trust in some way, their statements are more hedged than the above examples, mainly through the use of modals (e.g., \enquote{could}) or by speaking about \emph{appropriate} trust:

\begin{itemize} \label{cautious quotes}
    \item \enquote{[…] XAI will be key for both expert and non-expert users to enable them to have a deeper understanding and the appropriate level of trust […].} \cite{Clinciu2019Survey}
    \item \enquote{[…] comparative explanations could help establish a more appropriate level of trust.} \cite{Cai2019Effects}
    \item \enquote{[…] there is a need to explain […] so that users and decision makers can develop appropriate trust […].} \cite{Hoffman2018Explaining}
    \item \enquote{Explainable Machine Learning (XAI) […] enables human users to […] appropriately trust […] emerging generation of artificially intelligent partners.} \cite{Mathews2019Explainable}
    \item \enquote{[…] explanations are often proposed to […] moderate trust to an appropriate level […].} \cite{Wang2019Designing}
\end{itemize}

Overall, many authors assume some sort of systematic connection between trust and explanations. While some remain cautious about the exact nature of that relationship, many seem to endorse the straightforward relationship suggested by \ET.

\section{Empirical Evidence Concerning \ET}
\label{evidence}

Despite its intuitive appeal, \ET is not without problems. As we shall see in this section, the empirical evidence is not conclusive enough to support \ET.

\subsection{Empirical Findings}
\label{empirical findings}

Although there are empirical findings supporting the claim that explanations can lead to increased trust in systems \cite{Chakraborti2019Plan, Nagulendra2016Providing}, various empirical studies also provide evidence against that hypothesis. For instance, providing information about what kind of information will be analyzed within AI-based personnel selection can positively \emph{and} negatively affect variables that are commonly associated with trust towards intelligent systems (e.g., perceived fairness) \cite{Langer2018Information, Langer2021Spare, Newman2020Eliminating}. 

Furthermore, results by Schlicker et al. \cite{Schlicker2021Expect} indicate that providing an explanation does not affect healthcare professionals' perceived justice of automated scheduling decisions. Given that perceived justice is usually also associated with trust \cite{Colquitt2011Justice}, this finding provides further evidence against \ET.

These are just some of many examples where empirical research has found no support for the positive relation between explanations and trust (further examples are \cite{Chen2019User, Cheng2019Explaining}). In fact, some studies even found a negative effect of explanations on trust. For instance, Kizilcec et al. \cite{Kizilcec2016How} found that providing too much information eroded trust. Similarly, Papenmeier et al. \cite{Papenmeier2019How} found that the presence of an explanation either did not affect or even reduced trust. 

\subsection{Discussion of Empirical Findings}

Overall, there is some tension between previous empirical research and the various claims that explanations lead to trust. Thus, while it remains compatible with the data that some explanations will increase trust under certain conditions, \ET in its generality should not be assumed.
 
Once we take a closer look at the idea underlying \ET, these findings are not surprising. We can think of three straight\-forward reasons why explanations might fail to foster trust:

\begin{enumerate}
    \item If a person's trust in a system is already maximal, an explanation cannot further increase their trust.
    \item If the explanation reveals a problem of the system, the explanation might decrease rather than increase trust.
    \item If a person cannot comprehend the explanation or cannot use it to evaluate the system, the explanation might not change their trust in the system.
\end{enumerate}

Compelling arguments can be made that these reasons do indeed often play a role: Studies show that some people have a very high initial trust in automated systems \cite{Dzindolet2003Role}, explainability methods are often used for debugging systems \cite{Adadi2018Peeking, Carvalho2019Machine}, and many such methods produce explanations that are too technical for laypeople to understand \cite{Langer2021What, Gilpin2018Society}. It would be interesting to conduct research on whether these reasons are at play when explanations fail to increase trust. To this end, a meta-analysis could be a valuable starting point. For now, these considerations indicate why the relationship between explanations and trust is not as straightforward as assumed in \ET. Therefore, a requirement for explainability is not necessarily a suitable substitute for a requirement for trust in RE.

\section{From Trust to Trustworthiness}

Does the above discussion indicate that one should not try to engineer for trust via explainability? At this point, we can distinguish two motivations for why someone might want to elicit trust in a system: First, the developer or deployer of a system might want more people to use their technology. Second, we as a society might want reliable technologies that can improve our lives to receive the appropriate trust from their potential users and other stakeholders.

In the first case, the software developer or deployer might hope for trust independently of whether the system fulfills further desiderata like reliability, safety, or fairness. In other words, they might want users to trust their product whether or not it is actually trustworthy. In that case, explanations might not always help them reach their goal.

However, we can assume that many people who speak more generally about trust in technology, especially legislators, are interested in trust rather for the second reason. As we have seen in Section \ref{cautious quotes}, many of the more cautious quotes related to \ET focus on appropriate trust as opposed to trust in general. We will argue below that in the case where people are looking for the appropriate trust in a reliable system, explanations remain useful. An important mediator for such trust is a system's \emph{trustworthiness}, to which we will now turn.

\subsection{Differentiating Trust and Trustworthiness}

Trust is an attitude a stakeholder holds \emph{towards} a system. Trustworthiness, by contrast, is a property \emph{of} a system: intuitively, a system is trustworthy for a stakeholder when it is warranted for the stakeholder to put trust in the system.
While there are many different conceptualizations of trustworthiness \cite{McLeod2020Trust, hardin2002trust, hawley2019trustworthy, jones2012trustworthiness}, we will settle for an operationalization of trustworthiness that we deem suitable for the context of engineering artificial systems:

\begin{definition}[Trustworthiness] \label{def_TW}
A system $S$ is \emph{trustworthy} to a stakeholder $H$ in a context $C$ if and only if 
\begin{enumerate}
    \item[(a)] $S$ works properly in $C$, and
    \item[(b)] $H$ would be justified\footnote{We rely on an internalist notion of justification (cf.\ e.g., \cite{Pappas2017Internalist}).} to believe that (a) if $H$ came to believe that (a).
\end{enumerate}
\end{definition}

So, we see that trustworthiness is a property of a system that is parameterized with a stakeholder. Fulfilling condition (a) of Definition \ref{def_TW} is primarily up to the system, while fulfilling condition (b) also depends on the stakeholder in question.\footnote{In our view, the trustworthiness of a system can differ between stakeholders. For instance, a newly developed system for cancer detection might be trustworthy to its engineer who understands it in detail, but not to his friend, the oncologist, who does not have any insight into the system or any of its components.} Note that \enquote{works properly} is a deliberately vague expression. While it will be important to spell out this notion more precisely in future research on trust and trustworthiness, we shall not delve into the matter here. For current purposes, just note that merely fulfilling all specified requirements might not be enough for a system to \enquote*{work properly} in the sense of Definition \ref{def_TW}. An autonomous hiring system, for example, has to be just and fair in order to be considered as working properly, even if that has not been specified as an explicit requirement.

Ideally, we want both: that a given system is trustworthy \emph{and} that it is actually trusted. Unfortunately, though, the two can come apart. A judge might put great trust in a system that assesses defendants, while, in fact, the system might be racist and, therefore, not trustworthy. In this case, there is trust without trustworthiness, or \emph{unwarranted trust} \cite{jacovi2021formalizing}. Likewise, an elderly person, suspicious of new technological developments, may not trust their navigation system although they know that it works very reliably and will guide them to their destination safely and quickly. In this case, there is trustworthiness without actual trust, or \emph{failed trust} \cite{lee2004trust}. 

Looking back at the two potential reasons to engineer for trust we discussed above, it can be seen that trustworthiness is closely related to the idea of appropriate trust in a reliable system: The system's reliability is captured in part (a) of the definition above. Part (b) helps to ensure that if the person trusts the system, they are justified to do so and, thus, their trust is appropriate. Nevertheless, trustworthiness does not automatically guarantee the appropriate trust of all stakeholders.

\subsection{Trustworthiness as the Primary Concern}

If the system's trustworthiness does not necessarily go hand in hand with stakeholders' trust, the natural question to ask is which of the two should be given priority, even if we ideally want both. We argue that there are good reasons to give priority to trustworthiness.

\subsubsection{Practical Reasons}

From a pragmatic point of view, it is reasonable to spend less energy on features that designers can hardly control and instead prioritize whatever features are more controllable at design time \cite{Amershi2019Guidelines}. If we follow this reasoning, trustworthiness should take priority over trust, since our control over trust is very limited at design time, while we arguably have much better (though not complete) control over trustworthiness at design time. 

Recall that trustworthiness is mainly a property of the system, while trust is an attitude of the stakeholders. 
Granted, even trustworthiness is parameterized with a stakeholder, but this might be less troublesome than it initially looks:

Part (a) of Definition \ref{def_TW} is clearly controlled at design time, for it is the main objective of designers to make the system work properly, no matter how we spell this out. Part (b) seems more problematic, as it depends on specific stakeholders and what is justified for them to believe. This, however, is also not entirely outside the control of designers. In fact, designers have considerable control over (b) as they can already deliver appropriate justifications for certain stakeholders to believe in (a) as part of their system or alongside their system. (In the next section, we will see that explanations can be of help here.) 

Trust, on the other hand, can be controlled much less at design time: It can be elicited, for instance, by certain experiences a person has with a system, clever marketing and advertisement, or by the person's prior knowledge, beliefs, or preconceptions. So, whether someone trusts a system depends not only on its design and the stakeholders' interaction with it, but also heavily on the stakeholder's mindset, general attitude towards the system, prior experience with similar systems, and social network's attitude toward such systems \cite{Hoff2015Trust}. System designers can only influence some of these variables, while for others there is almost no possibility to influence them directly.

So, we can conclude that system designers have much less influence on the actual trust that people build in a system than the system's trustworthiness. Therefore, trustworthiness takes priority from a pragmatic point of view.

\subsubsection{Moral Reasons}

From a normative point of view, we may run a different argument coming to the same conclusion: If designers neglect trustworthiness and build an untrustworthy system, we will probably have either an untrustworthy system that most stakeholders will not trust in the long run or an untrustworthy system that is trusted mistakenly, which can have devastating consequences. Neither of these scenarios is desirable and, arguably, deploying a trustworthy system will frequently have morally better consequences, even if it is not trusted. Think back, for example, to the racist decision system in court. If an untrustworthy system is employed in court, it is much more likely to do wrong than a trustworthy system, regardless of whether it is trusted.

So, trustworthiness should often take priority, for even a trustworthy system that fails to spark trust can be expected to be morally superior to a similar untrustworthy system.

\subsubsection{Sustainability Reasons}

Trustworthiness may also prove to be the more sustainable desideratum compared to trust. An essential factor in a person's tendency to trust a system is the quality of experiences they have made with the system \cite{Bailey2007Automation, YuvilerGavish2011Effect, Manzey2012Human}. If people are convinced to trust a system that does not work properly, their trust might easily be violated if the system fails. Contrary to that, with a trustworthy system, people can adjust their level of trust to the system's abilities. Consequently, it will become less likely that the system disappoints people's expectations and, over time, a system that works very well will potentially gain more trust through positive experiences. 
 
Thus, while the stakeholders' trust in a system is also important, the system's trustworthiness is a worthy goal to engineer for and might even take priority before actual trust.

\subsection{Trustworthiness and Explainability}

Several authors have remarked upon the relation between trustworthiness and explanations \cite{Pierrard2019New, Friedrich2011Taxonomy, Polley2021Towards, Robbins2019Misdirected, Markus2021Role, Mittelstadt2019Explaining, Baum2017Challenges, Darlington2013Aspects,  McInerney2018Explore}. In a nutshell, their idea is that a system's explainability promotes its trustworthiness. If this idea holds up, it can serve as an important motivation behind XAI. Examples from the literature are:

\begin{itemize}
    \item \enquote{Explaining decisions […] by intelligent systems is […] essential for […] becoming trustworthy to humans.} \cite{Pierrard2019New}
    \item \enquote{[…] objectives of explanations are manifold, including aims such as increasing trustworthiness […].} \cite{Friedrich2011Taxonomy}
    \item \enquote{A trustworthy system should give fair and reliable results along with its explanations.}\cite{Polley2021Towards}
    \item \enquote{It should be clear that explicability is considered to be an important part of […] \enquote*{trustworthy} […] AI.} \cite{Robbins2019Misdirected}
    \item \enquote{[…] explainable AI can contribute to the bigger goal of creating trustworthy AI […].} \cite{Markus2021Role}
    \item \enquote{[…] xAI is to produce methods that make algorithmic decision-making systems more trustworthy […]} \cite{Mittelstadt2019Explaining}
    \item \enquote{To be ideally trustworthy, a […] system needs to provide us with a rationalizing explanation which is accurate, graspable, and permissible.} \cite{Baum2017Challenges}

\end{itemize}

We, too, claim such a connection: 
we suggested that designers have some control over the fulfillment of condition (b) of Definition \ref{def_TW}, namely by providing justification to the stakeholder to believe that the system \enquote*{works properly}. Plausibly, one way to do so is by giving explanations. The reasoning here is quite straightforward: if we want to be justified in our beliefs about how well a system works, it will often be helpful to have a sufficient understanding of the system. Accurate explanations can help us to gain this understanding and, therefore, the justification. So, while explanations might not help with trust, they are likely to help with trust\emph{worthiness}.

Note that this is not an empirical point but rather a theoretical one.  Granted, \emph{what} someone believes or whether they \emph{feel} justified in their beliefs are empirical questions of psychology. However, the question that we are after, namely whether someone's belief would be justified, is essentially a question of epistemology and, therefore, not an empirical one.

So, while we cannot assume \ET, our discussion suggests a tight connection between explanations and trustworthiness.

\section{Future Research Directions}

We argued that explainability can contribute to a system's trustworthiness and discussed why trustworthiness should often take precedence over trust in design processes. However, a range of questions remains to be answered by future research. For one thing, it is unclear how trustworthiness can be reliably assessed and measured. To this end, we need empirical and conceptual research to gain insights into what requirements to place on systems to make them trustworthy and how to meet these requirements. A more elaborate operationalization of system trustworthiness needs to be developed and agreed on and ways to assess trustworthiness have to be found.

A second issue that needs further research is spelling out the exact relationship between explainability and trustworthiness. It needs to be clarified which explanations, under which conditions, can justify a stakeholder's belief that the system works properly. With this in mind, we suggest paying particular attention to the context in which an explanation is given, as different stakeholders and situations might require different explanations to make the system trustworthy \cite{Langer2021What, Chazette2021Exploring}. 

Third, it remains important to investigate what role explanations can play to increase trust in a system. The findings we discussed in section \ref{empirical findings} indicate many unexplored factors in the relationship between explanations and trust that call for empirical research into this relationship. While it became evident that not \emph{all} explanations foster trust, there still is the strong suspicion that some explanations in the right contexts can actually do so -- and it remains to be seen which ones.  
To better understand how stakeholders build trust in a system based on explanations, it will, for example, be worth studying how the timing and presentation of explanatory information as well as stakeholders' expectations affect their trust-building. 

Fourth, future research should examine how to elicit, increase, and maintain stakeholders' trust in trustworthy AI systems. To this end, researchers should investigate how explainability and other (contextual) factors may work together and interact to determine trust. Work on this question may be closely tied up with research on the other issues just mentioned.

\section{Conclusion}

In summary, our exposition highlights three lessons for requirements engineers, developers, and researchers: first, current research 
does not imply a close relation between explanations or explainability and trust; second, trustworthiness is a property worth engineering towards; and third, further empirical research is needed to properly understand the relationship between explainability, trustworthiness, and trust.

These lessons have particular implications for RE: When designers want to ensure that stakeholders trust their system, they should not use explainability as a substitute -- at least according to current research. However, if they want to make their system trust\emph{worthy}, ensuring explainability might be very helpful and, thus, still of great importance. Also, one must not confuse trust and trustworthiness when formulating requirements. 
Overall, RE and many other disciplines would profit from more research on trust, trustworthiness, and explainability.

\section*{Acknowledgments}
Work on this paper was funded by the Volkswagen Foundation grants \textsc{AZ} 98509, 98512, 98513, and 98514 \href{https://explainable-intelligent.systems}{\enquote{Explainable Intelligent Systems}} (\textsc{EIS}) and by the \textsc{DFG} grant 389792660 as part of \href{https://perspicuous-computing.science}{\textsc{TRR}~248}. We thank three anonymous reviewers for their feedback.

{\raggedbottom
\bibliographystyle{IEEEtran}
\bibliography{bibliography}
}

\end{document}